\documentclass[%
reprint,
 amsmath,amssymb,
 aps,
]{revtex4-2}

\usepackage{graphicx}
\usepackage{dcolumn}
\usepackage{bm}
\usepackage{color}

\begin{document}

\preprint{APS/123-QED}

\title{Comment on “Tests and calibrations of nuclear track detectors (CR39) for operation in high neutron flux” by E.E. Kading \emph{et al.}, Phys. Rev. Res. 2, 023279 (2020)}


\author{Michael Paul} \email[Corresponding author: ]{paul@vms.huji.ac.il}
\affiliation{The Hebrew University of Jerusalem, Jerusalem 91904, Israel}

\author{Ulli Köster}
\affiliation{Institut Laue-Langevin, 71 avenue des Martyrs, 38042 Grenoble, France}

\date{\today}

\begin{abstract}
The conditions of tests and calibrations described in the article E.E Kading \emph{et al.} [Phys. Rev. Res. 2, 023279 (2020)] show striking flaws. 
The interpretation and extraction of quantitative results from this experiment are considered unreliable. 
All results of cross sections based on the results described in this article must be disregarded.

\end{abstract}

\maketitle


E.E.~Kading \emph{et al.} report in the paper \cite{kading2020} on the detection of alpha particles and protons using CR-39 solid state nuclear track detectors (SSNTD). 
Their results are the basis for the interpretation \cite{gai20,gai18} of the experimental data obtained in the framework of the project “Measurement of Neutron Interactions with $^{7}$Be and the Primordial $^{7}$Li Problem”
(funded by the US Israel Binational Science Foundation BSF 2012098), conducted at the Soreq Applied Research Accelerator Facility (SARAF).

We have already pointed out \cite{paul2019,schumann20} deficiencies in the preparatory studies and in the interpretation of the experiment mentioned above, preventing quantitative conclusions.
 In this comment, we itemize our objections as follows:

\begin{itemize}
\item 
	For etching the irradiated CR-39 samples, a simple glass beaker without thermostat was used. 
	The etching liquid was heated from below without stirring. The temperature profile inside was not recorded. 
\item
	The analysis of the depth profiles, which could have provided more reliable information about the energy of the detected particles, was not conducted. 
	Instead, only the radii of the induced pits were used for the evaluation. 
\item
	Only one run was used to generate each of the calibrations shown in Fig. 3 of \cite{kading2020}, although more experimental data were available. 
	Thus, the reproducibility of the calibrations is not assessed. 
	We have already stressed the particularly high uncertainty of the energy calibration of the proton response and the resulting unreliability of the reported cross section value for the $^{7}$Be(n,p) reaction \cite{schumann20}.
\item
The results of the experiments at SARAF and Institut Laue-Langevin (ILL) cannot be compared directly as claimed in Fig. 8 of \cite{kading2020}, because the irradiations were performed under drastically different conditions. In contrast to the SARAF setup (Fig. 4 of \cite{kading2020}), the CR-39 plate irradiated at ILL had been sandwiched between high-purity quartz/sapphire plates, sealed in Teflon bags, at atmospheric pressure and not in vacuum as claimed in \cite{kading2020}. The calibrations presented in \cite{kading2020} were performed in vacuum and are irrelevant to the ILL conditions in view of the strong dependence of track formation on pressure and environmental conditions during irradiation and processing (see \textit{e.g.} \cite{HER12,PRO94}).
\item
	Notice also that during the etching of the samples used for extracting the cross sections of the $^{7}$Be(n,$\alpha$) and $^{7}$Be(n,p) reactions, no calibration samples were simultaneously co-processed. 
	Moreover, the detector response to protons shown in Fig.3 bottom in \cite{kading2020} was analyzed with a different microscope and with different magnification than for the experimental runs in \cite{gai20,gai18}. 
	Consequently, all results reported in \cite{gai20,gai18} have been produced without valid calibration.
\end{itemize}

The experimental setup for the etching procedure used in \cite{kading2020} and the evaluation of the signals are far behind the present state-of-the-art performance for such detector systems (for comparison see for instance \cite{sinenian11, sinenian14}). 
We conclude that the uncertainties associated with the poorly controlled processing of the CR-39 detectors and the ana\-lysis of the experimental signals are not suited for a quantitative analysis. 
Therefore, it is our opinion that any cross section extracted or to be extracted from the experiment above performed in the project “Measurement of Neutron Interactions with $^{7}$Be and the Primordial $^{7}$Li Problem” using tests and calibrations in \cite{kading2020} must be disregarded.



\providecommand{\noopsort}[1]{}\providecommand{\singleletter}[1]{#1}%

\end{document}